\documentclass[11pt]{article}
\usepackage{graphics,color,amsfonts,amsthm,comment,epsf,epsfig,latexsym,url}

\advance \topmargin by -\headheight
\advance \topmargin by -\headsep
\evensidemargin \oddsidemargin
\def\proof{\par\noindent {\it Proof: \enspace}}

\def\ourskip{\vskip9pt}
\def\qed{\nolinebreak\hfill$\Box$\ourskip}

\def\mxth{\mathsurround=0pt}
\def\eqalign#1{\lineskip8pt\null\,\vcenter{\openup\jot\mxth
  \ialign{\strut\hfil$\displaystyle{##}$&$\displaystyle{{}##}$\hfil
      \crcr#1\crcr}}\,}

\newtheorem{definition}{Definition}
\newtheorem{theorem}{Theorem}

\newtheorem{proposition}{Proposition}
\newtheorem*{ellem}{Ellipse Lemma}
\newtheorem*{ellem1}{Ellipse Lemma (1)}
\newtheorem*{ellem2}{Ellipse Lemma (2)}
\newtheorem*{ellem3}{Ellipse Lemma (3)}
\newtheorem*{iellem}{Inverted Ellipse Lemma}

\newif\ifabstract
\abstracttrue
\newif\iffull
\ifabstract \fullfalse \else \fulltrue \fi

{\textsf{BEGIN ABSTRACT}#1\textsf{END ABSTRACT}}
\long\def\iffull#1\fi{\textsf{BEGIN FULL}#1\textsf{END FULL}}
\fi

\def\pargr#1{\par\ourskip\noindent\textbf{#1}}

\begin{document}

\def\R2{{\mathbb R}^2}
\def\ITLB{\mathrm{ITLB}}
\def\OPT{\mathrm{OPT}}

\let\epsilon=\varepsilon

\def\cp{{\sc closest-point}}
\def\fpa{{\sc farthest-point}}
\def\clptext{\hbox{closest-}\linebreak[0]\hbox{possible}}
\def\clp{\hbox{\rm closest-}\linebreak[0]\hbox{\rm possible}}
\def\fptext{\hbox{farthest-}\linebreak[0]\hbox{possible}}
\def\fp{\hbox{\rm farthest-}\linebreak[0]\hbox{\rm possible}}

\bibliographystyle{alpha}

\title{Optimal Adaptive Algorithms for
       Finding the Nearest and \\ Farthest Point
       on a Parametric Black-Box Curve}

\author{%
  Ilya Baran%
    \thanks{MIT Computer Science and Artificial Intelligence Laboratory,
            32 Vassar Street, Cambridge, MA 02139, USA,
            \{\texttt{ibaran}, \texttt{edemaine}\}\texttt{@mit.edu}}
\and
  Erik D. Demaine\footnotemark[1]%
}

\date{}

\maketitle
\begin{abstract}
We consider a general model for representing and manipulating
parametric curves, in which a curve is specified by a black box
mapping a parameter value between $0$ and $1$ to a point in Euclidean
$d$-space.  In this model, we consider the
\emph{nearest-point-on-curve} and \emph{farthest-point-on-curve}
problems: given a curve~$C$ and a point~$p$, find a point on $C$
nearest to~$p$ or farthest from~$p$.  In the general black-box model,
no algorithm can solve these problems.  Assuming a known bound on the
speed of the curve (a Lipschitz condition), the answer can be
estimated up to an additive error of $\epsilon$ using $O(1/\epsilon)$
samples, and this bound is tight in the worst case. However, many
instances can be solved with substantially fewer samples, and we give
algorithms that adapt to the inherent difficulty of the particular
instance, up to a logarithmic factor. More precisely, if
$\OPT(C,p,\epsilon)$ is the minimum number of samples of~$C$ that
every correct algorithm must perform to achieve tolerance $\epsilon$,
then our algorithm performs $O(\OPT(C,p,\epsilon) \log
(\epsilon^{-1}/\OPT(C,p,\epsilon)))$ samples.  Furthermore, any
algorithm requires $\Omega(k \log (\epsilon^{-1}/k))$ samples for some
instance $C'$ with $\OPT(C',p,\epsilon) = k$; except that, for the
nearest-point-on-curve problem when the distance between $C$ and $p$
is less than $\epsilon$, $\OPT$ is $1$ but the upper and lower bounds
on the number of samples are both $\Theta(1/\epsilon)$.  When bounds
on relative error are desired, we give algorithms that perform
$O(\OPT\cdot\log(2+(1+\epsilon^{-1})\cdot{}m^{-1}/\OPT))$
samples (where $m$ is the exact minimum or maximum distance from $p$ to
$C$) and prove that $\Omega(\OPT\cdot\log(1/\epsilon))$ samples are
necessary on some problem instances.
\end{abstract}

\section{Introduction}

Computational geometry has traditionally been focused on polygonal
objects made up of straight line segments.  In contrast, applications
of geometric algorithms to computer-aided design and computer graphics
usually involve more complex curves and surfaces.  In recent years,
this gap has received growing attention with algorithms for
manipulating more general curves and surfaces, such as circular arcs
\cite{Devillers-Fronville-Mourrain-Teillaud-2002}, conic arcs
\cite{Berberich-Eigenwillig-Hemmer-Hert-Mehlhorn-Schoemer-2002,
Wein-2002}, and quadratic surfaces \cite{Lennerz-Schoemer-2002}.  The
most general type of curve commonly considered in this algorithmic
body of work is a piecewise bounded-degree polynomial (algebraic)
curve, although such curves are not usually manipulated directly and
are more typically assumed to govern some process such as the motion
of a polygon in kinetic collision detection
\cite{Basch-Erickson-Guibas-Hershberger-Zhang-1999}.

\pargr{Parametric Black-Box Curves.}
A much more general model for specifying curves is the
\emph{parametric black-box model} that represents a curve
in Euclidean $d$-space as a
function $C\colon [0,1] \to \mathbb{R}^d$.  The only operation that
can be performed is to \emph{sample} (evaluate) the function at a
given parameter value $x \in [0,1]$.

Solving any nontrivial problem on a black-box curve requires some
additional conditions on the behavior of the curve.
We assume the
\emph{Lipschitz condition} that $\|C(x_1)-C(x_2)\| \leq L
|x_1-x_2|$ for all $x_1, x_2 \in [0,1]$, for a known constant~$L$.
Any piecewise-$C^1$ curve has such a
parameterization. By uniformly scaling the curve in
$\mathbb{R}^d$, we can assume that the Lipschitz constant $L$
is~$1$.

\begin{figure*}
    \begin{center}
    \input{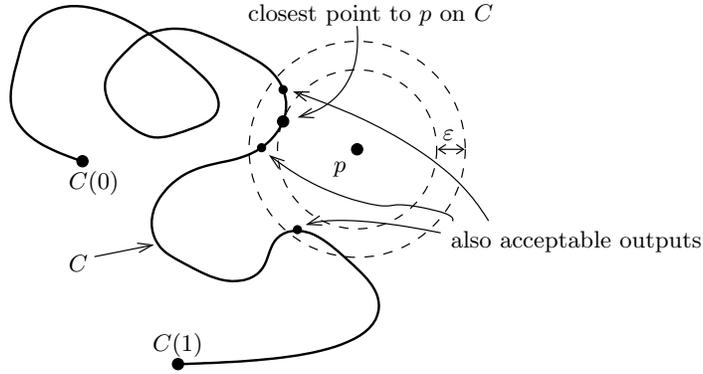}
    \caption{An instance of the nearest-point-on-curve problem.}
    \label{general}
    \end{center}
    \end{figure*}

\pargr{Nearest- and Farthest-Point-on-Curve Problems.}
In this paper, we solve two of the most basic proximity queries about
black-box Lipschitz curves: given a curve $C$ and a point $p$,
find a point on $C$ that is closest to~$p$ (\emph{nearest point}),
and find a point on~$C$ that is farthest from~$p$ (\emph{farthest point}).
In the black-box model, these problems are impossible to solve exactly,
because an algorithm will never, in general, sample
the nearest or farthest point.  Thus, a problem instance also specifies
an additive error tolerance $\epsilon$, and our goal is to find a point on
the curve $C$ whose distance to the point $p$ is within $\pm \epsilon$
of the minimum or maximum possible.
See Figure~\ref{general}.  Although we focus on absolute (additive)
error in this paper, we show in Section~\ref{relerror} how to modify
the absolute-error algorithms to obtain relative-error algorithms
(whose output is accurate to within a factor of $1+\epsilon$) that
have nearly optimal adaptive performance.

\pargr{Hard and Easy Instances.}
Any nearest-point-on-curve or farthest-point-on-curve instance can be solved
using $1/2\epsilon + O(1)$ samples: $C(0), C(2\epsilon), C(4\epsilon),
\dots, C(1)$. Unfortunately, this many samples can be necessary in
the worst case. For example, when $C(x) = q$ for all $x$ outside
an interval of length $2 \epsilon$ where at speed $1$ the curve moves
toward $p$ and then returns to~$q$, we need $1/2 \epsilon - O(1)$
samples to find the interval. Thus, worst-case analysis is not
very enlightening for this problem.

On the other hand, many instances are substantially easier.
As an extreme example, if $C$ is a unit-length line segment, then two samples,
at $C(0)$ and $C(1)$, completely determine the curve by the Lipschitz condition.

\pargr{Adaptive Analysis.}
Because the instance-specific optimal number of samples varies widely
from $\Theta(1)$ to $\Theta(1/\epsilon)$, we use the \emph{adaptive
analysis} framework, considered before in the context of boolean set
operations \cite{Demaine-Lopez-Ortiz-Munro-SODA-2000} as well as
sorting \cite{Estivill-Castro-Woods-1992} and aggregate ranking
\cite{Fagin-Lotem-Naor-2001}.  In the adaptive analysis framework, the
performance of an algorithm on a problem instance is compared to
$\OPT$, the performance of the best possible algorithm for that
specific problem instance.  By definition, for every problem instance,
there exists an algorithm that achieves $\OPT$ on that instance.  The
question is whether \emph{one} adaptive algorithm uses roughly
$\OPT(C,p,\epsilon)$ samples for \emph{every} instance
$(C,p,\epsilon)$.

\pargr{Our Results.}
We develop adaptive algorithms that solve the nearest-point-on-curve
and farthest-point-on-curve problems using
$O(\OPT(C,p,\epsilon)\log(\epsilon^{-1}/\OPT(C,p,\epsilon)))$ samples;
except that, for the nearest-point-on-curve problem when the distance
between $C$ and $p$ is less than~$\epsilon$, the number of samples may
be $\Theta(1/\epsilon)$, yet $\OPT = 1$.  We also prove that these
algorithms are \emph{optimally adaptive} in the sense that no adaptive
algorithm can achieve a strictly better bound (up to constant factors)
with respect to $\OPT$ and $\epsilon$.  Specifically, we show that,
for any $\epsilon > 0$ and $k > 0$, there is a family of curves $C$
each with $\OPT(C,p,\epsilon) = k$ such that every algorithm (even
randomized) requires $\Omega(k \log (\epsilon^{-1}/k))$ samples on
average for a curve $C$ selected uniformly from the family; and there
is a family of instances of the nearest-point-on-curve problem where
the distance between $C$ and $p$ is less than $\epsilon$ such that
every algorithm requires $\Omega(1/\epsilon)$ samples on average, but
$\OPT$ is 1.

\pargr{Related Work.}
Because our curve model is a black box, the problems that we consider
here have natural formulations in information-based complexity terms
(see \cite{Traub-Wasilkowski-Wozniakowski-1988} for an overview).
However, information-based complexity is primarily concerned with
worst-case, average-case, or randomized analysis of more difficult
problems, rather than adaptive analysis of algorithms for easier
problems as in this paper.  Information-based complexity does consider
adaptive algorithms (algorithms for which a query may depend on the
answers to previous queries), but primarily when they are more
powerful than non-adaptive algorithms in the worst (or average, etc.)
case, such as for binary search.

The problem of maximizing a Lipschitz function has been studied
in the context of global optimization.  This problem essentially
corresponds to the special case of the
nearest- or farthest-point-on-curve problem in
which $d=1$.  Beyond worst-case analysis, many algorithms for this
problem have been studied only experimentally
(see, e.g. \cite{Hansen-Jaumard-1995}),
but Piyavskii's algorithm
\cite{Piyavskii-1972}
has been previously
analyzed in what is essentially the adaptive framework, first in
\cite{Danilin-1971}.
The analysis was sharpened in 
\cite{Hansen-Jaumard-Lu-1991}
to show that
the number of samples the algorithm performs on $(C,\epsilon)$ is at
most 4 times $\OPT(C,\epsilon)$.  As Theorem~\ref{adaptivelower}
shows, this analysis cannot generalize to $d>1$.

Practitioners who manipulate curves and surfaces typically use
numerical algorithms, which are extremely general but sometimes
fail or perform poorly, or specialized algorithms for specific
types of curves, such as B-splines. Some algorithms for
manipulating general parametric curves and surfaces
guarantee correctness,
but the theoretical performance of
these algorithms is either not analyzed \cite{Snyder-1992}
or analyzed only in the worst-case \cite{Johnson-Cohen-1998, Guenther-Wong-1990}.
At the heart of our algorithm is G\"unther and Wong's \cite{Guenther-Wong-1990}
observation
that the portion of a Lipschitz curve between two nearby sample points
can be bounded by a small ellipse, as described in Section~\ref{Main Idea}.

\section{Problem Statement}

We use the real RAM model, which can store and manipulate exact real
numbers in $O(1)$ time and space.  Manipulation of real numbers
includes basic arithmetic ($+$, $-$, $\times$, $\div$), comparisons,
and $n$th roots.  We separately analyze the number of samples and the
additional computation time.
Although we describe and analyze our algorithms in $\R2$, both the
algorithms and their analyses trivially carry over to ${\mathbb R}^d$
for $d>2$.

We assume without loss of generality that the Lipschitz constant is 1,
that the parameter space is the unit interval, and that the query
point $p$ is the origin~$O$.  Throughout our discussion of
nearest-point-on-curve, $d_{\min}$ refers to the minimum distance from
$C$ to the origin.  Analogously, $d_{\max}$ denotes the maximum
distance from $C$ to the origin.  We assume that $\epsilon$ is smaller
than $1/2$ because otherwise, a single sample at $1/2$ immediately
solves both problems.  The two problems we consider are to find a
point on $C$ whose distance to $O$ is approximately $d_{\min}$ or
$d_{\max}$:

\ourskip\noindent {\bf Problem Nearest-Point-On-Curve~}
{\it Given a Lipschitz curve $C$ and an $0<\epsilon<1/2$, find a
parameter $x$ such that $\|C(x)\|\le d_{\min}+\epsilon$. }

\ourskip\noindent {\bf Problem Farthest-Point-On-Curve~}
{\it Given a Lipschitz curve $C$ and an $0<\epsilon<1/2$, find a
parameter $x$ such that $\|C(x)\|\ge d_{\max}-\epsilon$. }

\section{Nearest-Point-On-Curve: Adaptive Algorithm and its Analysis}

\begin{figure}
    \begin{center}
    \input{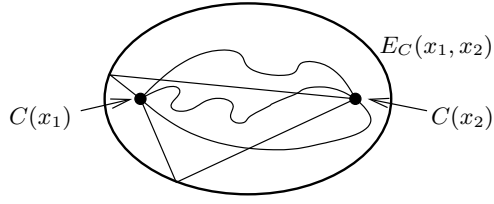}
    \caption{Some possible curves $C$ inside an ellipse.}
    \label{mainidea}
    \end{center}
    \end{figure}

\subsection{Main Idea}
\label{Main Idea}

The main observation is that, if we have sampled $C$ at $x_1$ and $x_2$,
then for $x$ between $x_1$ and $x_2$,
$\|C(x)-C(x_1)\|+\|C(x)-C(x_2)\| \le |x_2-x_1|$.  This means that when
the parameter $x$ is between $x_1$ and $x_2$, $C(x)$ stays within an
ellipse with foci at $C(x_1)$ and $C(x_2)$, whose major axis (sum of distances
to foci from a boundary point) has length $|x_2-x_1|$.  See
Figure~\ref{mainidea}.  Note that this ellipse is tight: by changing $C$
only between $x_1$ and $x_2$, we can force it to pass through any
point in the ellipse while keeping $C$ Lipschitz.  The following
propositions formalize this idea:

\begin{definition}
Given a Lipschitz curve $C$ and an interval
$[x_1,x_2] \subseteq [0,1]$, define the ellipse
$$E_C(x_1,x_2)= \Big\{p\in\R2 \;\Big|\;
\|C(x_1)-p\|+\|C(x_2)-p\|\le x_2-x_1\Big\}.$$
\end{definition}

\begin{proposition}
For an interval $J=[x_1,x_2] \subseteq [0,1]$, $C(J) \subseteq E_C(x_1,x_2)$.
\label{inellipse}
\end{proposition}
\proof Let $x \in J$.  By the Lipschitz condition,
$\|C(x_1)-C(x)\| \le x - x_1$ and similarly
$\|C(x_2)-C(x)\| \le x_2-x$. Adding these,
we get $\|C(x_1)-C(x)\|+\|C(x_2)-C(x)\|\le x_2-x_1$, so
$C(x) \in E_C(x_1,x_2)$. \qed

\begin{proposition}
\label{tightellipse}
Let $J=(x_1,x_2)\subseteq [0,1]$ and let $C$ be a
Lipschitz curve.  Then for every point $p$ in $E_C(x_1,x_2)$, there is
a Lipschitz curve $C'$ such that $C(x)=C'(x)$ for $x \not \in J$ and
for some $x \in J$, $C'(x) = p$.
\end{proposition}
\proof We can make $C'$ on $J$ consist of a line segment from $C(x_1)$
to $p$ and another one from $p$ to $C(x_2)$.  Because the total length
of these line segments is at most $x_2-x_1$, we can parametrize $C'$
at unit speed (or less) on $J$. \qed

The following proposition will often be used implicitly in our
reasoning:
\begin{proposition}
If $J'=[x_1',x_2']$ and $J=[x_1,x_2]$ and $J' \subseteq J$, then
$E_C(x_1',x_2') \subseteq E_C(x_1,x_2)$.
\end{proposition}
\proof
If $p\in E_C(x_1',x_2')$ then
$\|C(x_1')-p\|+\|C(x_2')-p\|\le x_2'-x_1'$ by definition.
We have $\|C(x_1)-C(x_1')\|\le x_1'-x_1$ and
$\|C(x_2)-C(x_2')\|\le x_2-x_2'$ by the Lipschitz condition on $C$.
Adding the three inequalities and applying the triangle inequality
twice, we get:
$$
\eqalign{
\|C(x_1)-p\|+\|p-C(x_2)\|&\le\cr\le
\|C(x_1)-C(x_1')\|&+\|C(x_1')-p\|+\|p-C(x_2')\|+\|C(x_2')-C(x_2)\|
\le x_2-x_1.
}
$$
So $p\in E_C(x_1,x_2)$, as required.
\qed

For notational convenience, let {\it $\clptext(x_1,x_2)$} denote the minimum
distance from a point in $E_C(x_1,x_2)$ to the origin.

\subsection{Proof Sets}
\label{Proof Sets}
The properties of $E_C$ immediately suggest a criterion for
determining whether a set of points on a curve is sufficient to
guarantee that a point sufficiently close to $O$ is among those in the
set: the distance from $O$ to the nearest
sampled point and the distance from $O$ to the nearest ellipse (around
adjacent points) should differ by at most $\epsilon$.

\begin{definition}
Let $P=\{x_1,x_2,\dots,x_n\}$ be a set of parameters in $[0,1]$
so that $0=x_1<x_2<\cdots<x_n=1$.  Let
$x_{\min} \in P$ be an element that minimizes $\|C(x_i)\|$.
Then $P$ is a \emph{proof set} if\/
$\|C(x_{\min})\|-\epsilon\le\clp(x_i,x_{i+1})$ for all~$i$.
\end{definition}

The following proposition shows that producing a proof set is the
only way an algorithm can guarantee correctness.

\begin{proposition}
\label{proofsamplesolves}
Let $P=\{x_1,x_2,\dots,x_n\} \subseteq [0,1]$ so that $0=x_1<x_2<\cdots<x_n=1$.
Let $x_{\min}$ be an element of $P$ that minimizes $\|C(x_i)\|$.
If $P$ is a proof set, then for any curve $C'$ such that
$C'(x_i)=C(x_i)$, $x_{\min}$ is a solution to
nearest-point-on-curve.
Conversely, if $P$ is not a proof set, there is a curve $C'$ such
that $C'(x_i)=C(x_i)$ for all $i$ and for which $x_{\min}$ is not
a solution.
\end{proposition}
\proof
For any curve $C'$ for which $C'(x_i)=C(x_i)$, $P$ is a proof set
for $C'$ precisely when it is a proof set for $C$.
Applying Proposition~\ref{inellipse},
we find that $C'([0,1])$ is contained in the union
of the ellipses $E_C(x_i,x_{i+1})$.  So, if $P$ is a proof
set, $\|C'(x_{\min})\|-\epsilon \le \|C'(x)\|$ for all $x \in [0,1]$,
which implies that $x_{\min}$ is a solution for $C'$.

Conversely, if $P$ is not a proof set, then there is a point $p$ in
some ellipse $E_C(x_i,x_{i+1})$ such that
$\|C(x_{\min})\|-\epsilon>\|p\|$.  By Proposition~\ref{tightellipse},
we can construct a curve that coincides with $C$
except in $(x_i,x_{i+1})$ and passes through $p$.  For this curve,
$x_{\min}$ will not be a solution. \qed

The requirement that $x_1=0$ and $x_n=1$ allows the analysis to avoid
special cases.  An algorithm could guarantee correctness without
sampling these endpoints, but because this saves only a constant amount
of work, we ignore this possibility in favor of simpler analysis.

\subsection{Algorithm Description and Correctness}
As we sample the curve, we maintain a set of ellipses around the
unsampled intervals.  At each step, we take the interval whose ellipse
is closest to the origin and sample in the middle of it, thus
replacing it with two smaller intervals (with smaller ellipses).  When
the sampled points form a proof set, we terminate and output the
closest point of those sampled.

Let $Q$ be a priority queue that stores triples of real numbers
$(d,x_1,x_2)$ sorted by $d$.  The algorithm is as follows:

\def\mx{\hat x_{\min}}
\def\md{\hat d_{\min}}

\ourskip
\cp$(C,\epsilon)$\ourskip
\nopagebreak[4]
\noindent
\nopagebreak[4]
1. Add $(\clp(0,1), 0, 1)$ to $Q$ \\ \noindent
\nopagebreak[4]
2. If $\|C(0)\|<\|C(1)\|$ then $(\mx,\md)\gets(0,\|C(0)\|)$ \\
$\strut\quad\;$ else $(\mx,\md)\gets(1, \|C(1)\|)$ \\ \noindent
\nopagebreak[4]
3. Do until finished:\\ $\strut\quad\;$
\nopagebreak[4]
4. $(d, x_1, x_2) \gets$ {\sc extract-min}$(Q)$ \\ $\strut\quad\;$
\nopagebreak[4]
5. If $\md - \epsilon \le d$ then output $\mx$ and {\sc stop}\\ $\strut\quad\;$
\nopagebreak[4]
6. $x \gets (x_1+x_2)/2$ \\ $\strut\quad\;$
\nopagebreak[4]
7. If $\|C(x)\| < \md$ then $(\mx, \md) \gets (x, \|C(x)\|)$ \\ $\strut\quad\;$
\nopagebreak[4]
8. Add $(\clp(x_1,x), x_1,x)$ to $Q$ \\ $\strut\quad\;$
\nopagebreak[4]
9. Add $(\clp(x,x_2), x,x_2)$ to $Q$
\ourskip

Correctness follows from Proposition~\ref{proofsamplesolves}: the
algorithm stops when the points sampled form a proof set and outputs the
closest point.  To show termination, we note that no interval of
length $2\epsilon$ or less is ever subdivided:
\begin{proposition}
\label{termination}
If in line 5, $x_2-x_1 \le 2 \epsilon$, \cp\ terminates at this line.
\end{proposition}
\proof Because $\md$ stores the minimum known distance to a point,
$\md \le \|C(x_1)\|$ and $\md \le \|C(x_2)\|$.  Let $p$ be a
point in $E_C(x_1,x_2)$ such that $\|p\| = d$.  Then by the definition
of $E_C$, $\|C(x_1)-p\|+\|C(x_2)-p\| \le 2\epsilon$.  This means that
at least one of $\|C(x_1)-p\| \le \epsilon$ or
$\|C(x_2)-p\|\le\epsilon$ is true.  If $\|C(x_1)-p\| \le \epsilon$,
then, by the triangle inequality, $\|C(x_1)\|-\|p\|\le \epsilon$.
This implies that $\md-d \le \epsilon$ so the algorithm stops.
Similarly for the other possibility. \qed

From this proposition, we can conclude that \cp\ stops after at most
$O(1/\epsilon)$ loop iterations because only $O(1/\epsilon)$ sample points
at least $\epsilon$ apart can fit in $[0,1]$, and in each iteration of
the loop, the
algorithm always samples one new point in an interval of width at
least $2\epsilon$.

\subsection{Ellipse Lemma}

To analyze \cp, we will make use of one geometric fact in three
incarnations:

\begin{ellem}
Let $0 \le x_1 \le x_2 \le x_3 \le x_4 \le 1$.  Also, let
$d,a\in{\mathbb R}$ with $0 < a < d$.  If $\clp(x_1,x_2)\le d$,
$\clp(x_3,x_4)\le d$, and $\|C(x_2)\|\ge d+a$, then
$\clp(x_1,x_4)\le d-a$.
\end{ellem}
\begin{figure}
    \begin{center}
    \input{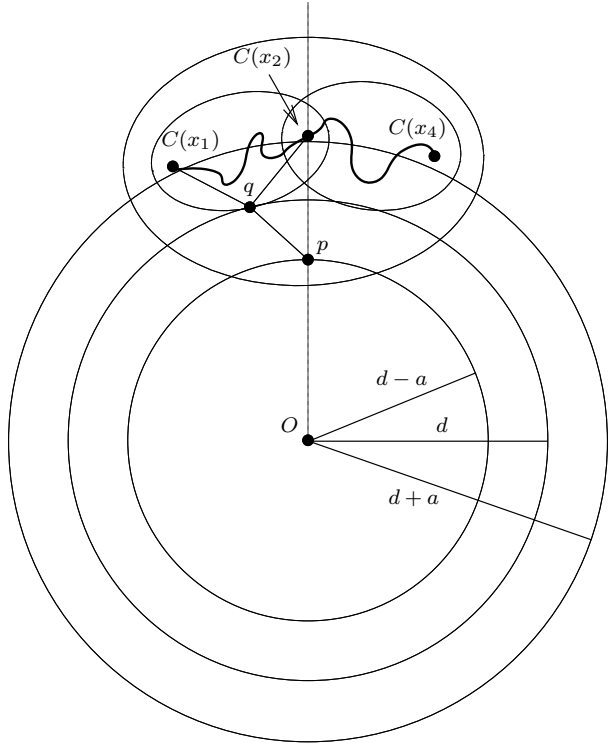}
    \caption{Ellipse Lemma}
    \label{ellipselemma}
    \end{center}
    \end{figure}
\proof
See Figure~\ref{ellipselemma}.
We may assume without loss of generality that $x_2=x_3$, because if
$\clp(x_3,x_4)\le d$, then $\clp(x_2,x_4) \le d$.  Let $p$ be the
intersection of the circle $\|v\|=d-a$ and the ray from the origin
through $C(x_2)$.  Obviously, $\|p\|= d-a$.  We will show that
$\|C(x_1)-p\|\le x_2-x_1$ and $\|C(x_4)-p\|\le x_4-x_2$.  This will
prove that $p \in E_C(x_1,x_4)$, and therefore $\clp(x_1,x_4)\le d-a$.

Because $\clp(x_1,x_2)\le d$, there is a point $q$ such that
$\|q\|\le d$ and $\|C(x_1)-q\|+\|q-C(x_2)\|\le x_2-x_1$.  We may set
the axes so that $C(x_2)$ is on the $y$ axis.  So let
$C(x_2)=(0, y)$.  Then $y\ge d+a$ and $p=(0, d-a)$.  Now if
$q=(x_q,y_q)$, then $\|q-C(x_2)\|=\sqrt{x_q^2+(y-y_q)^2}$ and
$\|q-p\|=\sqrt{x_q^2+(y_q-(d-a))^2}$.  Because $\|q\|\le d$, $y_q\le d$,
we have $(y_q-(d-a))^2\le((d+a)-y_q)^2\le(y-y_q)^2$, which means
that $\|q-p\| \le \|q-C(x_2)\|$.  Using this, the triangle inequality,
and the construction requirement of $q$, we get:
$$
\eqalign{
\|C(x_1)-p\|&\le \|C(x_1)-q\|+\|q-p\| \le\cr&\le
\|C(x_1)-q\|+\|q-C(x_2)\| \le x_2-x_1.
}
$$
The argument that $\|C(x_4)-p\|\le x_4-x_2$ is symmetric.
\qed

When generalizing this lemma from $\R2$ to $\mathbb{R}^d$, we consider
separately the planes through $O, C(x_1), C(x_2)$ and through
$O, C(x_2), C(x_4)$.

We will use the Ellipse Lemma in three different places in the
analysis, so we prove three simple corollaries:

\begin{ellem1}
Let $[x_1,x]$ and $[x,x_2]$ be intervals.  Let $0 < \epsilon < d$.
If $\clp(x_1,x_2) \ge d-\epsilon$ and
$\|C(x)\|\ge d$, then $\clp(x_1,x) \ge d-\epsilon/2$ or
$\clp(x,x_2)\ge d-\epsilon/2$ or both.
\end{ellem1}
\proof Suppose for contradiction that this is not the case: 
that we have
$\clp(x_1,x)<d-\epsilon/2$ and $\clp(x,x_2)<d-\epsilon/2$.
Let $d'$ be the larger of $\clp(x_1,x)$ and $\clp(x,x_2)$.
We have $d'<d-\epsilon/2$.  Now let $a=d-d'$ and apply
the Ellipse Lemma to $[x_1,x]$, $[x,x_2]$, $d'$ and $a$ to
get $\clp(x_1,x_2)\le d'-a$.  But
$d'-a=2d'-d<d-\epsilon$, which contradicts
the assumption that $\clp(x_1,x_2)\ge d-\epsilon$. \qed

\begin{ellem2}
Let $[x_1,x_2]$ and $[x_3,x_4]$ be intervals with $x_3\ge{}x_2$.
Let $0<\epsilon/2<d$.
If $\clp(x_1,x_2) \le d$, $\clp(x_3,x_4) \le d$, and
$\|C(x_2)\|>d+\epsilon/2$, then $\clp(x_1,x_4)<d-\epsilon/2$.
\end{ellem2}
\proof Let $a=\|C(x_2)\|-d$ so $a > \epsilon/2$.  Now apply
the Ellipse Lemma to $[x_1,x_2]$, $[x_3,x_4]$, $d$, and $a$ to
get $\clp(x_1,x_4)\le d-a < d-\epsilon/2$.
\qed

\begin{ellem3}
Let $[x_1,x]$ and $[x,x_2]$ be intervals and suppose $0 < \epsilon/2 < d$.
If $\clp(x_1,x_2) \ge d-\epsilon/2$ and
$\|C(x)\|>d+\epsilon/2$, then $\clp(x_1,x)>d$ or $\clp(x,x_2)>d$ or
both.
\end{ellem3}
\proof
Suppose that $\clp(x_1,x)\le d$ and $\clp(x,x_2) \le d$.  Apply the
Ellipse Lemma(2) to get that $\clp(x_1,x_2)<d-\epsilon/2$, which is a
contradiction.
\qed

\subsection{OPT}
We define the OPT of a problem
instance to be the number of samples that the best possible algorithm
makes on that instance.  In our analysis, we use the fact that
OPT is equal to the size of the smallest proof set, which
follows from Proposition~\ref{proofsamplesolves}.
Note that
OPT depends on
$C$ and $\epsilon$, but we write $\OPT$ or $\OPT(\epsilon)$ instead of
$\OPT(C,\epsilon)$ when
the arguments are clear.  For the analysis of \cp\ with
$\epsilon<d_{\min}$ we need the following estimate:
for any curve $C$, $\OPT(\epsilon/2)=O(\OPT(\epsilon))$.
We prove this by starting with
a proof set for $\epsilon$, inserting a new sample point in between
every pair of sample points in the proof set, and using the
Ellipse Lemma with a
continuity/connectedness argument to show that we can force the result
to be a proof set for $\epsilon/2$.

\begin{proposition}
\label{itlbgrowth}
If $\epsilon < d_{\min}$, for
any problem instance $(C,\epsilon)$, $\OPT(C,\epsilon/2) \le 2 \OPT(C,\epsilon)$.
\end{proposition}
\proof Consider a proof set $P$ of size $\OPT(\epsilon)$.
Let $x_i$
be the $i^{\rm th}$ smallest element of $P$.
Because $P$ is a proof set, $\clp(x_i,x_{i+1}) \ge d_{\min}-\epsilon$.
Now let
$$\eqalign{A&=\{x \in [x_i,x_{i+1}]\;|\;\clp(x_i,x) \ge d_{\min}-\epsilon/2\}
\cr
B&=\{x\hspace{-.6pt} \in \hspace{-.6pt} [x_i,x_{i+1}]\;|\;\clp(x,x_{i+1}) \ge d_{\min}-\epsilon/2\}} $$

By the Ellipse Lemma(1), $A \cup B = [x_i,x_{i+1}]$.  Also, because
$\clp(x_i,x_i)=\|x_i\| \ge d_{\min}$ and
$\clp(x_{i+1},x_{i+1})=\|x_{i+1}\|\ge d_{\min}$, $A \ne \emptyset$ and
$B \ne \emptyset$.  Because $\clp$ is continuous in both variables, $A$
is closed relative to $[x_i,x_{i+1}]$, being the preimage of the
closed set $\{t\;|\; t\ge d_{\min}-\epsilon\}$ under $t = \clp(x_i,x)$
with respect to the second variable.  Similarly, $B$ is closed
relative to $[x_i,x_{i+1}]$.  Because $[x_i,x_{i+1}]$ is connected,
$A\cap B\ne\emptyset$, so let $x \in A\cap B$.  This means
$\clp(x_i,x)\ge d_{\min}-\epsilon/2$ and
$\clp(x,x_{i+1})\ge d_{\min}-\epsilon/2$.  So for every pair of
adjacent samples in $P$, we can insert a new sample $x$ between them
($x$ may, of course, coincide with one of the samples already in $P$,
in which case we ignore it) so that in the resulting
set, $\clp(x_j,x_{j+1})\ge{}d_{\min}-\epsilon/2$ for all $j$.
Thus, we will have inserted at most
an additional $|P|-1$ elements.  In order to make the result a proof
set, we may need to insert one more element, $x$ such that
$\|C(x)\|=d_{\min}$.  This will make the result into a proof set
for $\epsilon/2$ with $2|P|$ elements.
\qed

\subsection{Phases}
We split an execution of \cp\ into two phases and analyze
each phase separately, giving an upper bound on the number of curve
samples.  The phases are a construction for the analysis only; the
algorithm does not know which phase it is in.  The
algorithm starts out in Phase~1, and switches to Phase~2 when all of
the ellipses around intervals stored in $Q$ are no closer than
$d_{\min}-\epsilon/2$ to the origin.
The distance from the ellipses in $Q$ to the origin can only
grow (as ellipses close to the origin are replaced by ellipses farther
away), so once the algorithm enters Phase~2, it can never leave it.
Let $P$ be a proof set for $\epsilon/2$ whose size is
$\OPT(\epsilon/2)$. We show that in each phase, the
number of samples is $O(|P| \log(\epsilon^{-1}/|P|))$.
We will want the following easy fact:

\begin{proposition}
\label{logsum}
Let $a_i$ for $1 \le i \le |P|$ be positive real numbers.  If we have
$\sum_{i=1}^{|P|} a_i \le \epsilon^{-1}$, then
$\sum_{i=1}^{|P|} \log a_i \le |P| \log(\epsilon^{-1}/|P|)$.
\end{proposition}
\proof
By the arithmetic-geometric mean inequality, we have
$\root |P| \of{\prod_{i=1}^{|P|} a_i} \le \frac{\sum_{i=1}^{|P|} a_i}{|P|}\le \frac{\epsilon^{-1}}{|P|}$.
Taking the logarithm of both sides gives us
$\frac{\sum_{i=1}^{|P|} \log a_i}{|P|} \le \log(\epsilon^{-1}/|P|)$.
Multiplying both sides by $|P|$ gives us the desired result.
\qed

\subsection{Phase 1}
\begin{proposition}
\label{musthavepoint}
If
$\clp(x_1,x_2)<d_{\min}-\epsilon/2$, then $P$ must have a point in
the open interval $(x_1,x_2)$.
\end{proposition}
\proof
For contradiction, suppose that $P$ has no point in $(x_1,x_2)$. This means
that $P$ has two consecutive points, $x_1'$ and $x_2'$, such that
$[x_1,x_2]\subseteq[x_1',x_2']$. So $E_C(x_1,x_2)\subseteq E_C(x_1',x_2')$
and therefore, $\clp(x_1,x_2)\ge \clp(x_1',x_2')$, which means
that $\clp(x_1',x_2')<d_{\min}-\epsilon/2$.  Hence, $P$ cannot
be a proof set for $\epsilon/2$.
\qed

Let $J=[x_1,x_2]\subseteq [0,1]$ be an interval that is subdivided in
Phase~1.  This implies that $\clp(x_1,x_2) < d_{\min}-\epsilon/2$ so
by Proposition~\ref{musthavepoint}, $P$ must have a point in
$(x_1,x_2)$.  This means that any interval that is subdivided in Phase~1
contains a point of $P$.

We need to count the samples in this phase.  We achieve this by classifying
every subdivision as either a ``split'' or a ``squeeze''.  A
subdivision is a {\it split} if both resulting intervals contain
points from $P$ and a {\it squeeze} if one of the resulting intervals
has no points from $P$.  Because the number of splits cannot be more
than $|P|-1$, we only need to count squeezes.  If $J$ is an interval
in $Q$ at some point in the execution of Phase~1, let $S(J)$ be the
number of squeezes that have happened to intervals containing $J$ and
let $L(J)$ be the length of $J$.  We want the following invariant:

\begin{proposition}
\label{phase1invar}
If at some point during Phase~1 of the algorithm, the intervals that
intersect $P$ are $J_1,J_2,\dots,J_k$, then
$\sum_{i=1}^k2^{S(J_i)} L(J_i) = 1$.
\end{proposition}
\proof We proceed by induction on the number of subdivisions.  At the
start of the execution of \cp, $S([0,1])=0$ and $L([0,1])=1$ so the base case
is clearly true.  Suppose an interval $J_i$ is split into $J_{i1}$ and
$J_{i2}$.  Because no new squeezes have occurred,
$S(J_{i1})=S(J_{i2})=S(J_i)$ and $L(J_{i1})=L(J_{i2})=L(J_i)/2$.  So
$2^{S(J_i)}L(J_i)=2^{S(J_{i1})}L(J_{i1})+2^{S(J_{i2})}L(J_{i2})$ and
the sum is not changed.  If the interval $J_i$ is squeezed into
$J_{i1}$, then $S(J_{i1})=1 + S(J_i)$ and $L(J_{i1})=L(J_i)/2$ so
$2^{S(J_i)}L(J_i) = 2^{S(J_{i1})}L(J_{i1})$ and the sum is not changed
in this case either. \qed

\begin{proposition}
\label{phase1runtime}
There are $O(|P|\log(\epsilon^{-1}/|P|))$ samples in Phase~1.
\end{proposition}
\proof
As noted above, we only need to count the squeezes.
By Proposition~\ref{phase1invar}, at the end of Phase~1, if
$J_1,\dots, J_k$ contain points of $P$, then
$\sum_{i=1}^k 2^{S(J_i)} L(J_i) = 1$.  But because no interval of length
$\epsilon$ or less ever appears, $L(J_i)>\epsilon$ so
$\sum_{i=1}^k2^{S(J_i)} < \epsilon^{-1}$ and $k\le|P|$.  Using
Proposition~\ref{logsum}, we get
$\sum_{i=1}^kS(J_i)\le k\log(\epsilon^{-1}/k)=O(|P|\log(\epsilon^{-1}/|P|))$.
 Every
squeeze increases $\sum_{i=1}^kS(J_i)$ by 1 and no operation ever decreases
it, so the number of squeezes is at most
$O(|P|\log(\epsilon^{-1}/|P|))$. \qed

\begin{figure*}
    \begin{center}
    \input{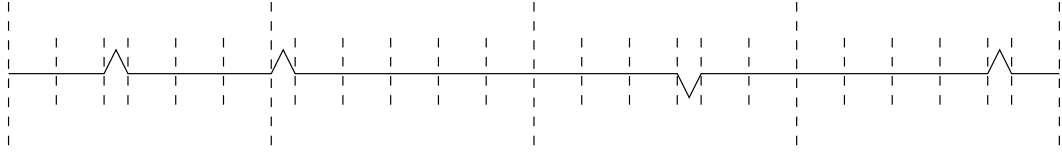}
    \caption{An example $C$ for the adaptive lower bound
for $n=24$ and $k=4$.}
    \label{lowerbound}
    \end{center}
    \end{figure*}

\subsection{Phase 2}

If in Phase~2 a point $x$ is sampled for which
$\|C(x)\|\le d_{\min}+\epsilon/2$, the algorithm stops.  Because we are
giving an upper bound on the running time, we may assume that every
point sampled is farther than $d_{\min}+\epsilon/2$ from the origin.

If $\clp(x_i,x_{i+1})>d_{\min}$, then $[x_i,x_{i+1}]$ will never be
chosen for subdivision.  This is because an interval around a point
that is $d_{\min}$ away from the origin has its ellipse at distance at
most $d_{\min}$ from the origin and will be chosen over
$[x_i,x_{i+1}]$.  Thus, let us call an interval $[x_i,x_{i+1}]$ {\em
alive} if $\clp(x_i,x_{i+1}) \le d_{\min}$ and call it {\em dead}
otherwise.  No dead interval is ever subdivided.

\begin{proposition}
\label{fewactive}
If $\epsilon < d_{\min}$, then when the \cp\ enters Phase~2, there are
$O(|P|)$ alive intervals.
\end{proposition}
\proof Let the alive intervals at the start of Phase~2 be
$[x_1,y_1],[x_2,y_2],\dots,[x_k,y_k]$.
From the assumption above, $\|C(x_i)\|>d_{\min}+\epsilon/2$ and
$\|C(y_i)\|>d_{\min}+\epsilon/2$.  Because the intervals are alive,
$\clp(x_i,y_i) \le d_{\min}$.  This means that we can apply the
Ellipse Lemma(2) to $[x_i,y_i]$ and $[x_{i+1},y_{i+1}]$ to get
$\clp(x_i,y_{i+1}) < d_{\min}-\epsilon/2$.  By
Proposition~\ref{musthavepoint}, $P$ has a point in $(x_i,y_{i+1})$.
Because at most two segments of the form $(x_i,y_{i+1})$ can overlap,
and each one has at least one point of $P$, there must be at most
$2|P|$ of these segments. \qed

Now suppose we subdivide an interval $[x_1,x_2]$ into $[x_1,x]$ and
$[x,x_2]$.  Because the algorithm is in Phase~2,
$\clp(x_1,x_2)\ge d_{\min}-\epsilon/2$.  By our assumption above,
$\|C(x)\|>d_{\min}+\epsilon/2$.  Applying the Ellipse Lemma(3), we
get that either $\clp(x_1,x)>d_{\min}$ or $\clp(x,x_2)>d_{\min}$
(or both).  This implies that when the interval is subdivided, at most
one of the resulting intervals can be alive.

\begin{proposition}
\label{phase2runtime}
If $\epsilon < d_{\min}$, then \cp\ performs at
most $O(|P|\log(\epsilon^{-1}/|P|))$
samples in Phase~2.
\end{proposition}
\proof
Let $l_1,l_2\dots,l_k$ be the lengths of the alive intervals at time
$t$.  Define $p(t)=\sum_{i=1}^k\log_2 (2l_i/\epsilon)$.  Because no
interval of length $2\epsilon$ or less is ever subdivided,
$l_i>\epsilon$ and so each term in the sum is at least 1.  At every
subdivision, an alive interval is replaced with at most one alive
interval of half the length; therefore, each subdivision decreases
$p(t)$ by at least 1.  This implies that the total number of
subdivisions cannot be greater than $p(t_0)$ where $t_0$ is the
time when the algorithm enters Phase~2.  Now consider the situation
at time $t_0$. The
total length of the alive intervals is at most 1, so we have
$\sum_{i=1}^k2l_i/\epsilon \le 2\epsilon^{-1}$.  Applying
Proposition~\ref{logsum} to this inequality and to the
definition of $p(t_0)$, we get
$p(t_0)\le k\log(2\epsilon^{-1}/k)$.
By Proposition~\ref{fewactive}, $k=O(|P|)$, so we get
$p(t_0)=O(|P|\log(\epsilon^{-1}/|P|))$, which means there are at most
that many samples of $C$ in phase~2. \qed

\subsection{Analysis Conclusion}
\begin{theorem}
\label{cpadaptive}
If on a problem instance with $\epsilon<d_{\min}$, we let
$n=\OPT(\epsilon)\log({\epsilon^{-1}}/{\OPT(\epsilon)})$,
algorithm \cp\ uses $O(n)$ samples and $O(n\log n)$ additional time,
where the constant in the $O$ notation is independent of the instance.
\end{theorem}
\proof
Combining Propositions~\ref{phase1runtime}, \ref{phase2runtime}, and
\ref{itlbgrowth}, we get that the number of samples the algorithm
makes is $O(n)$.  Because the samples are stored in a priority queue,
which may be implemented as a heap, it takes $O(\log n)$ time to
insert or extract a sample.  Hence the algorithm uses $O(n\log n)$ time
for the heap operations.
\qed

This theorem does not hold for $d_{\min}\le\epsilon$ because then the
condition $a<d$ would not be satisfied when we invoke the Ellipse Lemma.
The best conclusion we can make about the running time of \cp\ when
$d_{\min}\le\epsilon$ is that the number of samples is
$O(1/\epsilon)$.  Below we prove that it is impossible to do
better with respect to OPT.

\section{Lower Bounds}
\subsection{Worst-Case Lower Bound}
As mentioned in the introduction, in the worst case, we cannot do better than
the trivial algorithm:

\begin{theorem}
For any $\epsilon>0$, there is a problem instance of
nearest-point-on-curve on which any algorithm requires
$\Omega(\epsilon^{-1})$ samples.
\end{theorem}
\proof
Suppose we are given $\epsilon$.  Let $C$ be the constant ``curve'',
$C(x)=p$ for all $x\in [0,1]$ with $\|p\| > \epsilon$.  Now, for any
interval $[x_1,x_2]\subseteq [0,1]$, $E_C(x_1,x_2)$ is a circle centered at
$p$ whose radius is $(x_2-x_1)/2$.  This means that in any
proof set, every two points are less than $2\epsilon^{-1}$ apart in
the parameter space, so the OPT for this problem is
$\Theta(\epsilon^{-1})$. \qed

\subsection{Adaptive Lower Bound}

We prove that \cp\ is optimal with respect to the number of
samples of $C$.

\begin{theorem}
\label{adaptivelower}
For any algorithm, and for any $k\in\mathbb{N}$,
and any $\epsilon\in(0,1/k)$,
there is a problem
instance with $\OPT=O(k)$ on which that algorithm requires
$\Omega(k\log(\epsilon^{-1}/k))$ samples.
\end{theorem}
\proof
Let $\epsilon$ and $k$ be given.  
We will construct a
problem instance family for which
$k=\Omega(\OPT)$ and the number of
samples required by any algorithm is $\Omega(k\log (\epsilon^{-1}/k))$
on at least one instance of that family.
\par
Let $n=\epsilon^{-1}/3$.  Divide the parameter space into $n$ equal
regions and group them into $k$ groups of $n/k$ regions each.  In each group,
let the curve have one spike in some region
(and be flat in the other regions of that group).  Let $k-1$ of the
spikes point up, and let the remaining spike point down.
See Figure~\ref{lowerbound}.
The origin is far
below the curve and $\epsilon$ is less than the height of a
spike, so that the only solutions to a nearest-point-on-curve
instance of this form are on the
spike pointing down.  
Because an omniscient adversary may force the last spike the algorithm
examines to be the one pointing down, and the algorithm cannot
determine whether a spike points up or down without sampling on it,
the algorithm must find every spike.  Note that if $x$ is a point in
parametric space that corresponds to the boundary between groups,
$C(x)$ does not depend on where the spikes are chosen.  Moreover,
sampling inside one of the $k$ groups (and not on a spike) only gives
information about whether the spike in that group is to the left or to
the right of the point sampled.  This implies that the algorithm must
perform a binary search on each of the $k$ groups.  The minimum number
of samples to do this is indeed $\Omega(k\log(n/k))$.

To show that $k=\Omega(\OPT)$, we note that because the curve is
piecewise-linear (and parametrized at unit speed), placing a
point at every corner gives a proof set for any $\epsilon$, because
that completely determines the curve.  Each spike has 3 corners
and there are possibly two more endpoints, so $\OPT\le 3k+2$.
\qed

\subsection{Lower Bound for $d_{\min} \le \epsilon$}

\begin{theorem}
For any algorithm and for any $\epsilon>0$, there is a problem instance
with $d_{\min}\le\epsilon$ such that the algorithm requires
$\Omega(\OPT(\epsilon)\epsilon^{-1})$ samples to solve it.
\end{theorem}
\proof Because $d_{\min}\le\epsilon$, $\OPT(\epsilon)=1$.
To define $C$, let us split $[0,1]$ into $\epsilon^{-1}/4$ intervals of
width $4\epsilon$.  Fix one of these intervals,
$J=(x_1,x_1+4\epsilon)$.  For $x\not\in J$, let
$C(x)=(0,2.5\epsilon)$.  For $x\in J$ let
$$
C(x)=
\cases{
(0,2.5\epsilon-(x-x_1))\quad &
\hbox{for $x<x_1+2\epsilon$}\cr
(0,2.5\epsilon-(x_1+4\epsilon-x))\quad&\hbox{for $x\ge x_1+2\epsilon$}.
}
$$
Informally, one of the intervals has a spike of height $2\epsilon$
pointing at the origin.  Now, at $x=x_1+2\epsilon$,
$C(x)=(0,\epsilon/2)$, so $d_{\min}=\epsilon/2$ and $\OPT=1$.  The
only valid outputs on such a problem instance are points on the spike.
Because sampling the curve anywhere except $J$ gives no information on
the location of the spike, which could be in any of $\epsilon^{-1}/4$
possible intervals, an algorithm is forced to do a linear search that
requires $\Omega(\epsilon^{-1})$ samples.
\qed

These lower bounds also work for randomized algorithms,
because the reductions are from linear search and binary search,
problems for which randomized algorithms can do no better than
deterministic algorithms (up to constant factors).

\section{Farthest-Point-on-Curve}

\begin{figure}
    \begin{center}
    \input{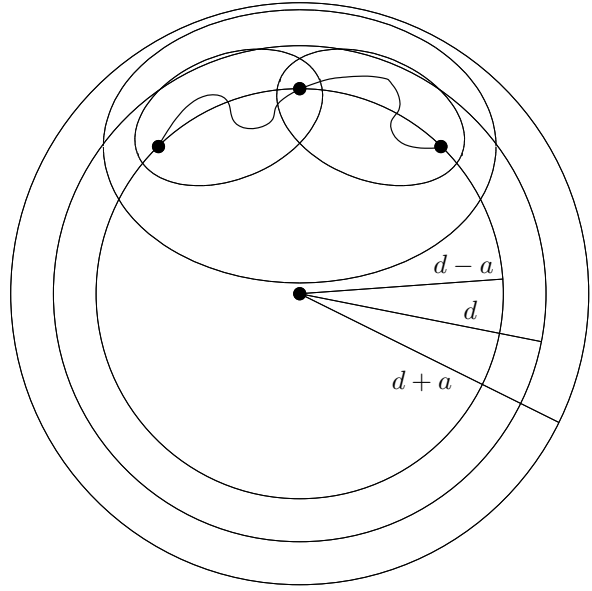}
    \caption{A counterexample to the inverted Ellipse Lemma
(ellipses are to scale)}
    \label{ellipsecounter}
    \end{center}
    \end{figure}

It is natural to consider the symmetric problem of finding a point on
$C$ whose distance to a given point is within $\epsilon$ of the
largest possible.  It is straightforward to modify \cp\ to \fpa, which
solves the farthest-point-on-curve problem.  The first two lower bounds for
nearest-point-on-curve hold for farthest-point-on-curve as well.  The
analysis is also easy to carry over to \fpa, with one exception: the
natural ``inversion'' of the Ellipse Lemma is false.
Figure~\ref{ellipsecounter} illustrates this.
 Nevertheless, the algorithm
running time is the same (to within a constant factor) because we can
prove a modified inverted Ellipse Lemma.  Note that $\fp(x_1,x_2)$
(the analogue of $\clp(x_1,x_2)$) refers to the maximum distance from
a point in $E_C(x_1,x_2)$ to the origin.  
To simplify the proof, we
impose an extra condition that $\|C(x_i)\|\le d-a$, which was
required only for $i=2$ in the original lemma.

\begin{iellem}
Let $0 \le x_1 \le x_2 \le x_3 \le x_4 \le 1$.  Also, let
$d,a\in{\mathbb R}$ such that $0 < a < d$.  If $\fp(x_1,x_2)\ge d$,
$\fp(x_3,x_4)\ge d$, and $\|C(x_i)\|\le d-a$ for $i\in\{1,2,3,4\}$, then
$\fp(x_1,x_4)\ge d+\frac{3}{5}a$.
\end{iellem}

First, we state and prove what is essentially a special case of this
lemma and then, to prove the lemma, we will reduce the general case to
this special case.

\begin{proposition}
\label{farellipsepro}
Let $a>0$, let $A$ and $B$ be points such that $\|A\|=\|B\|=1$ and let
$Q$ be the intersection of the bisector of angle
$AOB$ with the circle $\|v\|=1+a$.  Let $P$
be the intersection of ray $OB$ with circle
$\|v\|=1+8a/5$.  Then $\|A-Q\|+\|Q-B\| \ge \|A-P\|$.
\end{proposition}
\begin{figure}
    \begin{center}
    \input{farellipseprop.pstex_t}
    \caption{Illustration for proof of Proposition~\ref{farellipsepro}}
    \label{farellipseprop}
    \end{center}
    \end{figure}

\proof
See Figure~\ref{farellipseprop}.
First, note that $\|A-Q\|=\|Q-B\|$ so we need to show that
$2\|A-Q\|\ge\|A-P\|$.  Let $c$ be the cosine of angle $BOQ$.  Then the
cosine of the angle $BOA$ is $2c^2-1$ by the double angle formula.
Using the Law of Cosines, we write:
$$
\eqalign{
\|A-Q\|&=\sqrt{1+(1+a)^2-2(1+a)c}\cr
\|A-P\|&=\sqrt{1+\left(1+\frac{8a}{5}\right)^2-
2\left(1+\frac{8a}{5}\right)(2c^2-1)}.
}
$$
We prove the Proposition by expanding
$(2\|A-Q\|)^2-\|A-P\|^2$ and showing that it is
positive:
$$
4\|A-Q\|^2-\|A-P\|^2=
4+4(1+a)^2-8(1+a)c-1-\left(1+\frac{8a}{5}\right)^2+
2(2c^2-1)\left(1+\frac{8a}{5}\right),
$$
which simplifies to
\begin{equation}
\label{fareq}
4(c-1)^2+\frac{36a^2}{25}+\frac{8a}{5}(c-1)(4c-1)
\end{equation}

Notice that (\ref{fareq}) is quadratic in $c$ and the $c^2$ term has a
positive coefficient.  This means that it has a single global minimum
as $c$ varies.  Because we are trying to prove that (\ref{fareq}) is
positive, it is sufficient to show that the expression takes on a
positive value when $c$ is at the minimum.  To find the minimum, we
differentiate (\ref{fareq}) with respect to $c$ to get

$$
\frac{64a+40}{5}c-8a-8,
$$
which is equal to 0 for $c=\frac{5+5a}{5+8a}$.  Substituting it back
into (\ref{fareq}), we get
$$
\frac{36a^2}{(5+8a)^2}+\frac{36a^2}{25}-\frac{24a^2(15+12a)}{5(5+8a)^2}
=\frac{288a^3}{25(5+8a)}\ge 0,
$$
as desired.
\qed

\noindent{\it Proof of Inverted Ellipse Lemma:}
As in the original Ellipse Lemma, assume without loss of generality
that $x_2=x_3$.  Let us also assume that the condition on the points
is tight (it is straightforward to reduce to this case using the
triangle inequality), that is, $\|C(x_1)\|=\|C(x_2)\|=\|C(x_4)\|=d-a$.
Once again, let $P$ be the intersection of the circle
$\|v\|=d+\frac{3}{5}a$ and the ray from the origin through $C(x_2)$.
We need to show that $|x_2-x_1|\ge \|C(x_1)-P\|$ and the symmetric
case ($|x_4-x_2|\ge\|C(x_4)-P\|$) will follow.  From combining them,
we will be able to conclude the lemma.  When applying
Proposition~\ref{farellipsepro}, by scaling the entire picture, we can assume
that $d-a=1$ and so $d=1+a$.

In general, we only need to consider the case when the ellipse
$E_C(x_1,x_2)$ is as small as possible, that is, it is tangent to the
circle $\|v\|=d$.  Let $Q$ be a point of tangency.  Then (from the
optical properties of ellipses) the origin $O$ is on the bisector of
angle $C(x_1),Q,C(x_2)$.  Therefore, if we reflect the
line $Q,C(x_2)$ off the line $Q O$, $C(x_1)$ must be on
an intersection of the reflected line and the circle $\|v\|=d-a$.  Let
$D$ be the other intersection.  Because the reflection of $C(x_2)$ is
both on the circle $\|v\|=d-a$ and on the reflected line, it could be
either $C(x_1)$ or $D$.  If it is $C(x_1)$, we can use
Proposition~\ref{farellipsepro}, with $A\gets C(x_1)$ and $B\gets C(x_2)$ to
conclude that $\|C(x_1)-Q\|+\|Q-C(x_2)\|\ge\|C(x_1)-P\|$.  Otherwise,
$D$ is the reflection of $C(x_2)$ across $Q O$ and there are
two possibilities:

1. If $D$ is between $C(x_1)$ and $Q$, then we apply
Proposition~\ref{farellipsepro} with $A\gets D$ and $B\gets C(x_2)$ to get
$\|D-Q\|+\|Q-C(x_2)\|\ge\|D-P\|$.  We add
$\|C(x_1)-D\|$ to both sides and apply the triangle inequality
to get
$$
\eqalign{
|x_2-x_1| &\ge
\|C(x_1)-Q\| + \|Q-C(x_2)\|=
\|C(x_1)-D\| + \|D-Q\| + \|Q-C(x_2)\|\ge \cr &\ge
\|C(x_1)-D\|+\|D-P\|
\ge
\|C(x_1)-P\|.
}
$$

\begin{figure}
    \begin{center}
    \input{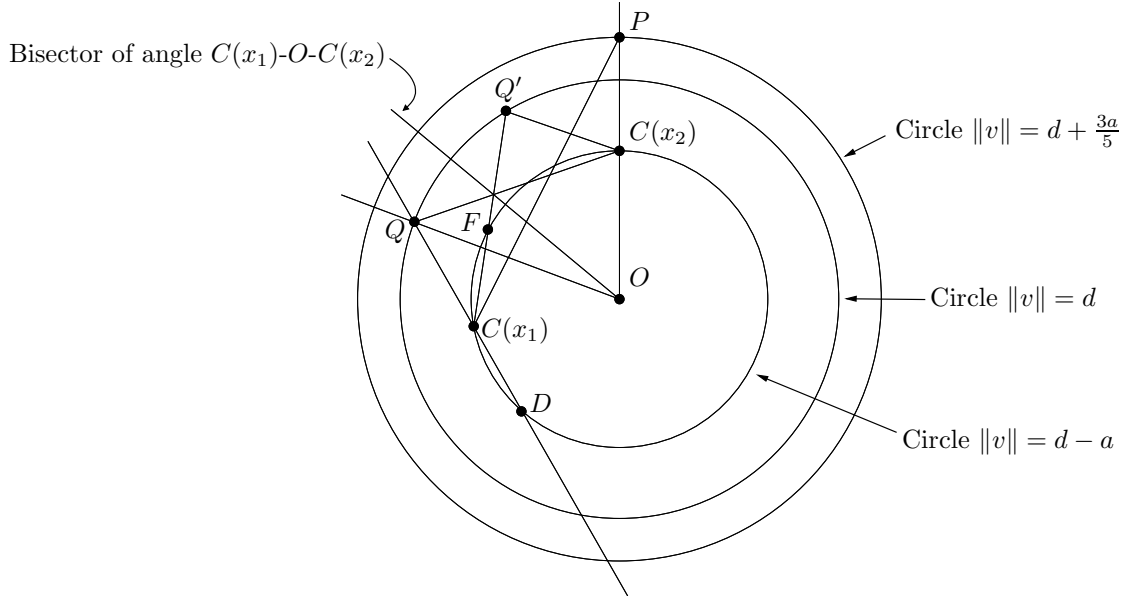}
    \caption{Illustration for the second case of the proof}
    \label{farellipse}
    \end{center}
    \end{figure}

2. Otherwise, $C(x_1)$ is between $D$ and $Q$.  See
Figure~\ref{farellipse}.  Let $Q'$ be the reflection of $Q$ across the
bisector of angle $C(x_1),O,C(x_2)$.  Then,
$\|C(x_1)-Q'\|=\|C(x_2)-Q\|$ and $\|C(x_2)-Q'\|=\|C(x_1)-Q\|$.  Now
the line segment $C(x_1),Q'$ has a second intersection with the
circle $\|v\|=d-a$ (besides the endpoint $C(x_1)$) because this line
segment is the result of reflecting line segment $D Q$ (which
contains $C(x_1)$, by assumption) first across $O Q$ and then
across the bisector of $C(x_1),O,C(x_2)$.  We will call
this intersection $F$.  $F$ is the image of $C(x_1)$ under the two
reflections described above, which implies that
$\|C(x_1)-Q\|=\|F-Q'\|$.  This means that $\|F-Q'\|=\|Q'-C(x_2)\|$ so
we can apply Proposition~\ref{farellipsepro} with $A\gets F$,
$B\gets C(x_2)$, and $Q\gets Q'$ to get
$\|F-Q'\|+\|Q'-C(x_2)\|\ge\|F-P\|$.  We add $\|C(x_1)-F\|$ to both
sides to get
$$
\eqalign{
|x_2-x_1| &\ge
\|C(x_1)-Q'\| + \|Q'-C(x_2)\|=
\|C(x_1)-F\| + \|F-Q'\| + \|Q'-C(x_2)\|\ge \cr &\ge
\|C(x_1)-F\|+\|F-P\|
\ge
\|C(x_1)-P\|,
}
$$
as desired.
\qed

Because the Inverted Ellipse Lemma has a weaker conclusion, in terms
of the constant, than the original Ellipse Lemma, the analogue of
Proposition~\ref{itlbgrowth} based on the Inverted Ellipse Lemma
states that $\OPT(5\epsilon/8)\le2\OPT(\epsilon)$.  This means that
in order to get that $\OPT(3\epsilon/8)=O(\OPT(\epsilon))$, which we
need for the analysis of Phase~2, we need to apply the analogue of
Proposition~\ref{itlbgrowth} three times (because $(5/8)^3<3/8$).

The analysis of farthest-point-on-curve does not have the problem that
nearest-point-on-curve has when $d_{\min}\le\epsilon$.  Every time the
Inverted Ellipse Lemma is used in the transformed proof, the condition
that $a>d$ holds regardless of the curve or $\epsilon$, unlike in
nearest-point-on-curve.

\begin{theorem}
On farthest-point-on-curve problem instance $(C,\epsilon)$, let
$n=\OPT(\epsilon)\log({\epsilon^{-1}}/{\OPT(\epsilon)})$.
Then algorithm \fpa\ uses
$O(n)$ samples and $O(n\log n)$ additional time.
\end{theorem}

\section{Relative Error}
\label{relerror}
We now examine modifications to our problems in which the goal is to
guarantee a relative error bound instead of an absolute error bound.
Specifically, for the nearest-point-on-curve problem, the objective
is a parameter $x$ such that
$\|C(x)\|\le(1+\epsilon)d_{\min}$; and for farthest-point-on-curve, we
need $\|C(x)\|\ge{}d_{\max}/(1+\epsilon)$.  We require that a
nearest-point (farthest-point) problem instance has $d_{\min}$
($d_{\max}$) nonzero, because otherwise the problem is unsolvable.
It turns out that simple modifications to the absolute-error algorithms
analyzed above yield adaptive relative-error algorithms.
For proving an upper bound on the number of samples used
by the algorithms, we focus on the nearest-point problem; for
farthest-point, the upper bound analysis is analogous.

We start by defining a proof set for a relative-error nearest-point
problem instance.  Let $P$ be a set of samples of $C$ that includes 0
and 1, let $U_P$ be the distance from the nearest point of $P$ to the
origin, and $L_P$ be the distance from the nearest ellipse around
adjacent points of $P$ to the origin.  We say that $P$ is a proof set
for the relative-error problem instance $(C,\epsilon)$ if $L_P>0$ and
$U_P/L_P\le1+\epsilon$.  It is easy to show the analogue to
Proposition~\ref{proofsamplesolves}, that a proof set for relative
error is required for a relative-error algorithm to guarantee
correctness.  So relative-error $\OPT$ is the size of a smallest proof
set, minus at most $2$ to account for the fact that including 0 and 1
may not be necessary.

To modify the absolute-error algorithm \cp, first note that as it executes,
$\hat{d}_{\min}$ is an upper bound on $d_{\min}$, and the top element
of $Q$ is a lower bound on $d_{\min}$.  Let us call these values
$U$ and $L$, respectively.  The termination condition in
line~5 is that $U-L\le\epsilon$.  If we replace it by the
condition that $L>0$ (to prevent division by zero) and
$U/L\le1+\epsilon$, we get a relative-error algorithm.

\sloppy
\begin{theorem}
The modified algorithm for the relative-error nearest-point-on-curve
problem uses
$O(\OPT\cdot\log(2+(1+\epsilon^{-1})\cdot{}d_{\min}^{-1}/\OPT))$
samples.
\end{theorem}
\fussy
\def\eabs{\epsilon_{\hbox{\tiny\sc ABS}}}
\proof
Let
$\eabs=\frac{\epsilon\cdot{}d_{\min}}{1+\epsilon}$.  Notice that if
$U-L\le\eabs$, then because
$L\le{}d_{\min}\le{}U$,
$$\frac{U}{L}\le
\frac{U}{U-\eabs}\le
\frac{d_{\min}}{d_{\min}-\eabs}
=\frac{1}{1-\frac{\epsilon}{1+\epsilon}}=1+\epsilon.
$$
So the relative-error algorithm with error $\epsilon$ terminates no
later than a hypothetical execution of the absolute-error algorithm
would with error $\eabs$.  By Theorem~\ref{cpadaptive}, we know that
such an absolute-error algorithm terminates after at most
$O(\OPT_{\rm{}ABS}\cdot\log(2+\eabs^{-1}/\OPT_{\rm{}ABS}))$ samples,
where $\OPT_{\rm ABS}$ is the absolute-error $\OPT$ for $\eabs$.  We
now have an upper bound on the running time of the modified
relative-error algorithm in terms of $\OPT_{\rm ABS}$.  To complete the proof,
we need to show a lower bound on $\OPT$ in terms of $\OPT_{\rm ABS}$.

In a relative-error proof set $P$,
$L_P\ge{}d_{\min}-\frac{\epsilon\cdot{}d_{\min}}{1+\epsilon}$, because
otherwise,
$$
\frac{U_P}{L_P}\ge
\frac{d_{\min}}{L_P}>
\frac{d_{\min}}{d_{\min}-\frac{\epsilon\cdot{}d_{\min}}{1+\epsilon}}=
\frac{1}{1-\frac{\epsilon}{1+\epsilon}}=1+\epsilon.
$$
So if we take a proof set for relative error $\epsilon$ and add a
sample at distance $d_{\min}$ from the origin, we obtain a proof set
for absolute error $\eabs$.  This proves that
$\OPT(\epsilon)+1\ge\OPT_{\rm{}ABS}$.  On the other hand, if we
have an absolute-error proof set $P$ for $\eabs$, we have
$U_P-L_P\le\eabs$, so $U_P/L_P\le(1+\epsilon)$, which implies that it
is also a relative-error proof set for $\epsilon$, and so
$\OPT(\epsilon)\le\OPT_{\rm{}ABS}$.  Therefore, the relative-error
algorithm performs
$O\left(\OPT(\epsilon)\cdot\log\left(2+\frac{1+\epsilon}{\epsilon\cdot{}d_{\min}}\Big/\OPT(\epsilon)\right)\right)$ samples.
\qed

We modify the construction used in proving Theorem~\ref{adaptivelower}
to prove a lower bound for the relative-error problem.

\begin{figure}
    \begin{center}
    \input{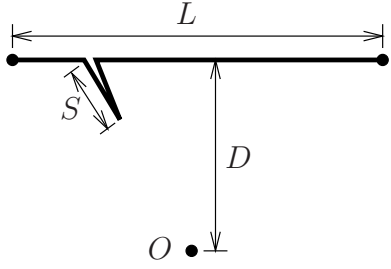}
    \caption{An example curve segment on which the proof of Theorem~\ref{adaptiverellower} is based}
    \label{lowerrel}
    \end{center}
    \end{figure}

\begin{theorem}
\label{adaptiverellower}
For any algorithm and for any $0<\epsilon<1$ and $k\in\mathbb{N}$, there
is a problem instance with $\OPT=O(k)$ on which that algorithm
requires $\Omega(k\log(\epsilon^{-1}))$ samples to solve the
relative-error problem.
\end{theorem}
\proof
Consider a piecewise-linear curve segment as shown in
Figure~\ref{lowerrel}.
Because such a segment is piecewise-linear, 5 samples are sufficient
to obtain all information about it.  We would like to show that for
some combinations of $S$, $L$, and $D$, the only solutions to the
relative-error problem are on the spike and it takes logarithmic time
to find it.

In order for the only solutions to the relative-error nearest-point
problem to be on the spike, the distance from $O$ to the tip of
the spike has to be smaller than $D/(1+\epsilon)$.  The distance from
the tip of the spike to $O$ is maximized when the spike is at one
of the endpoints of the curve segment.  In this case, the distance
from the tip of the spike to $O$ is $\sqrt{D^2+(L/2)^2}-S$.
So we need $D>(1+\epsilon)(\sqrt{D^2+L^2/4}-S)$, which is
equivalent to $S/L>\sqrt{(D/L)^2+1/4}-D/(L+L\epsilon)$.
If $D/L=\frac{1}{2\sqrt{\epsilon(2+\epsilon)}}$, the inequality
becomes:
$$
\frac{S}{L}>\sqrt{\frac{1}{4\epsilon^2+8\epsilon}+\frac{1}{4}}-
\frac{1}{(1+\epsilon)(2\sqrt{\epsilon^2+2\epsilon})}=
\frac{\sqrt{\epsilon^2+2\epsilon}}{2\epsilon+2}.
$$
So if we choose $S=L\sqrt{\epsilon}$, the above inequality is
satisfied (because
$(2\epsilon+2)\sqrt{\epsilon}=\sqrt{4\epsilon^3+8\epsilon^2+4\epsilon}>\sqrt{\epsilon^2+2\epsilon}$).

Therefore, we can construct a curve segment of arbitrarily small
length $L+2S$ with the only solutions to the nearest-point problem on
a spike of size $2S$, which is no more than $2 L\sqrt{\epsilon}$.
Sampling on the curve segment but not on the spike only gives
information whether the spike is to the left or to the right of the
point sampled.  Therefore, a binary search taking
$\Omega(\log((L+2S)/S))=\Omega(\log(1/\epsilon))$ steps is necessary
to find the spike.

To construct the curve, simply paste $k$ copies of curve segments, as
described above (they may overlap), except make $k-1$ of the spikes
point away from the origin and only one point toward it.  Because the
length of each curve segment can be arbitrarily small, the total
length can be made exactly 1 (and therefore, appropriately
parameterized, is a valid input).  The only solutions are on the spike
pointing toward the origin.  As in the argument for
Theorem~\ref{adaptivelower}, a binary search is required to find each
spike and a linear search on the curve segments is required to find
the spike pointing toward the origin, giving a lower bound of
$\Omega(k\log(1/\epsilon))$.  On the other hand, $\OPT\le5k$.

For farthest-point, the construction is analogous to the one above,
but ``flipped''.  On the curve segment on which the solution is
located, the spike points away from the origin.  To ensure that the
only solutions are on the spike, the distance from the tip of the
spike to $O$ has to be at least $\sqrt{D^2+(L/2)^2}$.  This distance
is minimized when the spike is in the middle of the curve segment and
the distance from the tip to $O$ is $D+S$.  Thus, we need
$D+S>(1+\epsilon)\sqrt{D^2+L^2/4}$, which is the same as
$S/L>(1+\epsilon)\sqrt{(D/L)^2+1/4}-D/L$.  Notice that the right hand
side is simply $(1+\epsilon)$ times the right hand side of the
analogous inequality for nearest-point.  Therefore, if $D/L$ is as for
nearest point and $S=L(1+\epsilon)\sqrt{\epsilon}$ (which is still
$O(L\sqrt{\epsilon})$), the inequality is satisfied and a binary
search on each curve segment requires $\Omega(\log(1/\epsilon))$
samples.  \qed

The upper and lower bounds for the relative-error problem do not
match.  We leave open the problem of finding an optimally adaptive
algorithm in this setting.


\section{Conclusion}
The results in this paper give asymptotically tight bounds on the
absolute-error nearest-point-on-curve and farthest-point-on-curve
problems in the adaptive framework.  We also show almost tight
bounds in the relative-error setting.  We believe that a similar
analysis can provide insight into the adaptive performance of
algorithms for other curve problems based on
Proposition~\ref{inellipse}, including those described in
\cite{Guenther-Wong-1990}.  We plan to carry out this analysis in the
future.  A more difficult open problem is generalizing
Proposition~\ref{inellipse} from one-dimensional curves to
two-dimensional surfaces in a way that allows algorithms based on the
generalization.

\bibliography{adaptive,curves,kinetic,glopt}

\end{document}